\documentclass[newstyle,onecolumn,proceedings]{rmaa}

\renewcommand{\P}[1]{%
\ifnum#1=1\hbox{OW~168--326E}\fi
\ifnum#1=2\hbox{OW~167--317}\fi
\ifnum#1=3\hbox{OW~163--317}\fi
\ifnum#1=5\hbox{OW~158--323}\fi
\ifnum#1=0\hbox{OW~171--334}\fi}

\def\hii{\protect{\sc H\thinspace ii}}
\def\neiii{[Ne {\sc iii}]}
\def\oiii{[O~{\sc iii}]}
\def\oii{[O~{\sc ii}]}
\def\siii{[S {\sc iii}]}
\def\sii{[S {\sc ii}]}
\def\siv{[S {\sc iv}]}

\def\sp3{S$^{+3}$}
\def\etap{$\eta^\prime$}
\def\etal{et\thinspace al.~}

\def\Hb{H$\beta$}

\def\hb{\rm H\beta}

\def\Zsol{{\rm\,Z_\odot}}

\def\tstar{$T_\star$}

\def\lam{$\lambda$}
\def\cc{{\rm\,cm^{-3}}}

\def\spose#1{\hbox to 0pt{#1\hss}}
\def\lta{\mathrel{\spose{\lower 3pt\hbox{$\mathchar"218$}}
     \raise 2.0pt\hbox{$\mathchar"13C$}}}
\def\gta{\mathrel{\spose{\lower 3pt\hbox{$\mathchar"218$}}
     \raise 2.0pt\hbox{$\mathchar"13E$}}}

\title{Calibrating nebular diagnostics of \tstar\ and abundance}

\author{M. S. Oey\altaffilmark{1,2}
\affil{Space Telescope Science Institute}
J. C. Shields
\affil{Ohio University}
M. A. Dopita
\affil{Australian National University}
R. C. Smith
\affil{Cerro Tololo Inter-American Observatory}}

\altaffiltext{1}{STScI Institute Fellow}
\altaffiltext{2}{Present address:  Lowell Observatory, 1400 W. Mars Hill Rd.
	Flagstaff, AZ   86001, USA (oey@lowell.edu)}

\fulladdresses{
\item M. S. Oey:
Space Telescope Science Institute, 3700 San Martin Drive, 
	Baltimore, MD   21218, USA (oey@stsci.edu)
\item J. C. Shields:
Ohio University, Dept. of Physics and Astronomy, Clippinger
	Research Labs. 251B, Athens, OH   45701, USA 
	(shields@helios.phy.ohiou.edu)
\item M. A. Dopita:
Research School of Astronomy and Astrophysics, 
	Australian National University, Private Bag,
	Weston Creek P.O., ACT 2611, Australia (Michael.Dopita@mso.anu.edu.au)
\item R. C. Smith:
Cerro Tololo Inter-American Observatory, Casilla 603, La
	Serena, Chile (csmith@noao.edu) }

\shortauthor{Oey et al.}
\shorttitle{Nebular emission-line diagnostics}

\keywords{galaxies: abundances --- \hii\ regions --- stars:
fundamental parameters --- Magellanic Clouds --- supernova remnants}

\abstract{%
We obtained nebular spectroscopy of LMC \hii\ regions having
classified stellar populations, thereby strongly constraining the
ionization input parameters.  Using photoionization models, we then
evaluate the performance of nebular diagnostics of \tstar\ and abundance.
We introduce \neiii/\Hb\ as a nebular diagnostic of the ionizing
stellar \tstar.  In contrast to the widely-used \etap\ parameter,
\neiii/\Hb\ has greater sensitivity to mid and early O-stars, and is
robust to nebular morphology and the presence of shocks.  We present a
preliminary calibration of both \tstar\ diagnostics for LMC
metallicity.  We also
introduce $S234\equiv$ (\sii\ + \siii\ + \siv)/\Hb\ as a diagnostic of S
abundance.  $S234$ is much less sensitive to the nebular ionization
parameter than is $S23$ or $R23$.  The intensity of \siv 10.5$\mu$ is
easily estimated from the optical and near-IR line ratios. 
We present calibrations of $S23$ and $S234$ that are reliable at
metallicities $Z \lta 0.5 \Zsol$.
}

\listofauthors{W.~J.~Henney \& S.~J.~Arthur}
\indexauthor{Henney, W.~J.}
\indexauthor{Arthur, S.~J.}

\begin{document}

\maketitle

\section{Introduction}

Nebular emission-line diagnostics are one of our most important probes
of conditions in extragalactic star-forming regions.  In particular,
these diagnostics are a primary means of constraining ionizing
stellar populations and metallicities in distant galaxies.  
However,
we rely heavily on photoionization models to interpret observed nebular
emission-line ratios, and comparisons between theoretical models and
observations have been lacking, especially for ordinary \hii\
regions.  Therefore, we have obtained long-slit spectroscopic
observations of four \hii\ regions in the Large Magellanic Cloud (LMC), for
the purpose of comparing the observed emission-line diagnostics with
predictions from photoionization models.  The sample spans a range in
stellar spectral type from O7 to WNE, and in nebular morphology from
classical Str\"omgren sphere to extreme shell.  Two objects
show evidence of shock excitation.  

We obtained both stationary, spatially-resolved observations of these
nebulae, and scanned, spatially-integrated observations.  A complete
investigation of the entire dataset is presented by Oey {\etal}(2000;
Paper~I) and Oey \& Shields (2000; Paper~II).  The
spatially resolved observations generally showed excellent agreement
between photoionization models and observations.  However, a puzzling
exception was our finding that the temperature-sensitive ratio
\oiii\lam4363/\lam5007 is {\it over}predicted by the models (Paper~I).
This is opposite to the expected effect of electron temperature
fluctuations (Peimbert 1967).  Our investigation of the abundance
determinations shows that it is essential to use an appropriate
relation between the electron temperatures in the high and low
ionization zones, which could otherwise introduce errors in the
abundance determinations of up to 0.2 dex.  Interestingly, we find no
evidence of abundance variations in DEM L199, a large \hii\ complex
dominated by three early-type WR stars.  For further details on our
results from the spatially resolved data, we refer the reader to
Papers I and II.  Below, we summarize results from the
spatially-integrated observations, which resemble those that would be
obtained if the LMC were at a distance of 15--20 Mpc.

\section{Diagnostics of \tstar}
\subsection{The \etap\ parameter}

\begin{figure}
\includegraphics[width=\textwidth]{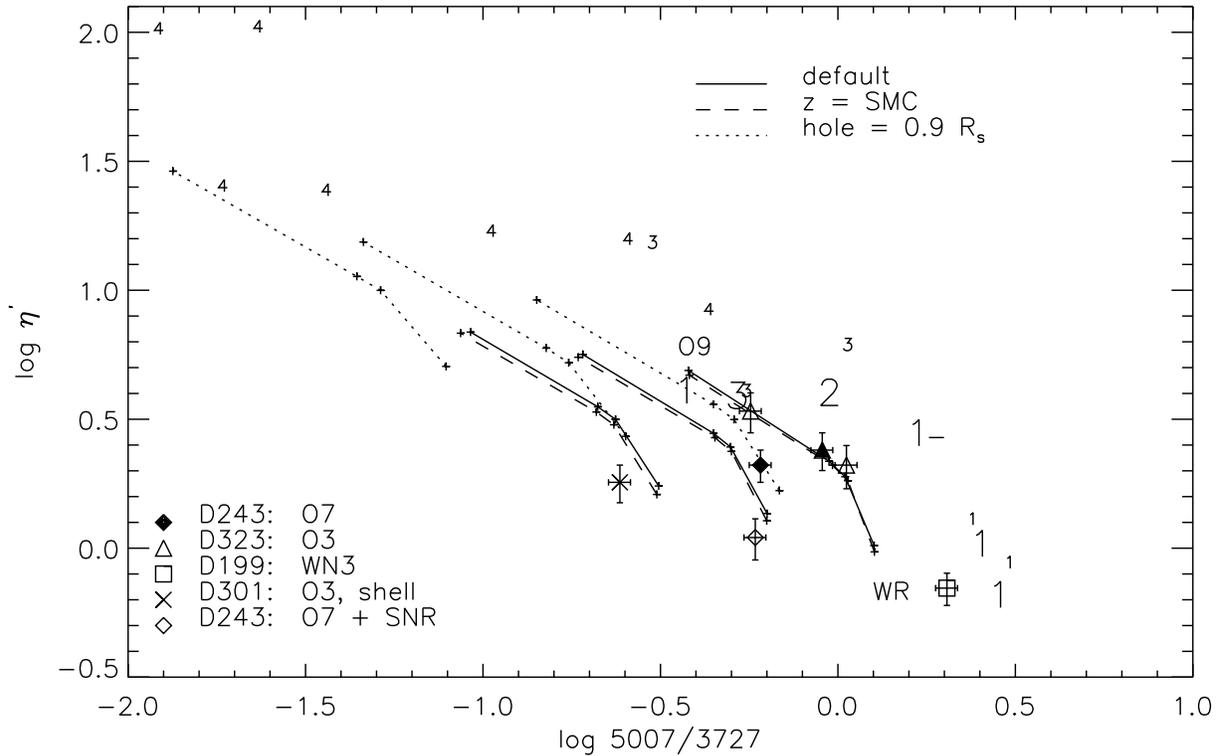}
\vspace*{-4.7in}
\caption{Model photoionization tracks for \etap\ vs \oiii/\oii,
using WR, O3, O7, and O9 stellar atmospheres.  Symbols show our data,
and numerals show nebular data from 
the Kennicutt {\etal}(2000) sample, with ionizing stellar spectral
types binned according to the key; objects along the bottom are
non-detections.  See text for more details.
\label{figetap}
}
\end{figure}

The most widely-used diagnostic of \tstar\ has been the
``radiation softness parameter'' of V\'\i lchez \& Pagel (1988),
defined as the relative ratios of singly to doubly-ionized O and S.
\begin{equation}\label{etap}
\eta^\prime \equiv \rm \frac{[O\thinspace II]\lambda3727/
	[O\thinspace III]\lambda\lambda4959,5007}
	{[S\thinspace II]\lambda6724/
	[S\thinspace III]\lambda\lambda9069,9532}
 \quad .
\end{equation}
Figure~\ref{figetap} shows theoretical tracks of \etap\
vs. \oiii/\oii, with our spatially integrated observations overplotted.
The models are generated with the photoionization code {\sc Mappings~II}
(Sutherland \& Dopita 1993), using stellar atmospheres from Schaerer \&
de Koter (1997) and Schmutz {\etal}(1992).  The default models (solid
lines) have LMC metallicity and central hole size of 0.1 times the Str\"omgren
radius ($R_{\rm s}$).  The three tracks have electron density varying
between 1, 10, and 100 $\cc$, which are directly proportional to
variations in the ionization parameter $U$, with increasing $U$
corresponding to increasing log \oiii/\oii.  The individual models
represent spectral types of roughly O9, O7, O3, and early WR, toward
decreasing values of \etap, respectively.  We also show model tracks
for a shell morphology having an inner radius of 0.9 $R_{\rm s}$
(dotted lines), and tracks having SMC metallicity but otherwise the
same as the default.  The data points for our sample of LMC nebulae are
plotted with symbols identified in the key.  Numerals show objects 
from the Kennicutt {\etal}(2000) sample, with large and small numerals
indicating LMC and Galactic objects, respectively; these data are not
spatially integrated over the entire nebulae (see Paper~I).

For both the models and data, Figure~\ref{figetap} shows that \etap\
can effectively discern the presence of WR stars and late-type O stars.
However,  mid- and early-type O stars yield highly degenerate values
of \etap.  Thus, this diagnostic is essentially insensitive to
$T_\star \gta 40,000$ K, with the exception of the WR stars.
In addition, Figure~\ref{figetap} shows that \etap\ is also sensitive
to nebular morphology.  The model tracks for a hollow shell strucure
(dotted lines) are clearly offset from the default models.

Peimbert {\etal}(1991) showed that the inclusion of supernova remnants
(SNRs) within integrated \hii\ region spectra can affect interpretation
of the nebular conditions.  The object DEM 243 encompasses s SNR, and
Figure~\ref{figetap} shows data points for the \hii\ region both with
and without the SNR included (open and solid diamonds, respectively).
DEM 301 (cross) also shows evidence of shock 
excitation (Paper~I).  It is apparent in Figure~\ref{figetap} that the
\etap\ values of DEM 243 with the SNR, and DEM 301, overestimate
\tstar\ in view of the actual stellar spectral types.  This therefore
shows that \etap\ is also sensitive to the presence of shocks.

\subsection{The \neiii/\Hb\ parameter}

To address these shortcomings in the \etap\ diagnostic,
we therefore introduce \neiii/\Hb\ as a complementary diagnostic of \tstar.
Although this ratio is sensitive to abundance, the high ionization
potential (40.96 eV) for Ne~{\sc iii} yields greater sensitivity to
hotter ionizing spectral types.  Likewise, this high ionization
potential reduces sensitivity to shock excitation.

\begin{figure}
\includegraphics[width=\textwidth]{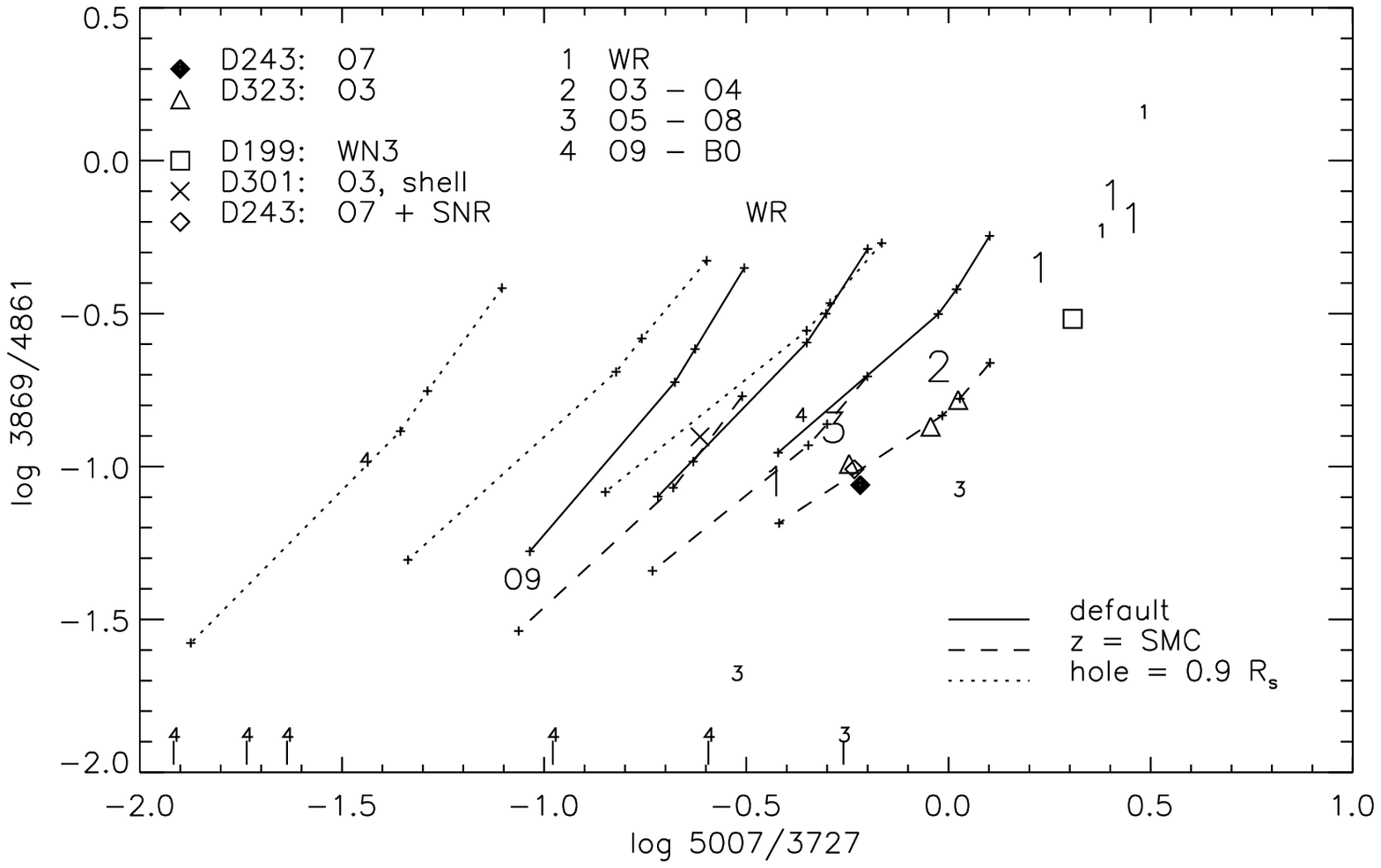}
\vspace*{-4.7in}
\caption{Same as Figure~\ref{figetap} for \neiii/\Hb\ vs \oiii/\oii.
\label{figneiii}
}
\end{figure}

In Figure~\ref{figneiii} we show model tracks of log \neiii/\Hb\ vs log
\oiii/\oii, for sequences in \tstar, with symbols as in Figure~\ref{figetap}.
Figure~\ref{figneiii} shows that \neiii/\Hb\ is a fairly successful
discriminant of \tstar, even between mid and early O-types.  The
dotted lines also show that it is highly insensitive to nebular
morphology.  In addition, the two data points for DEM 243 including
and excluding the SNR (solid and open diamonds, respectively)  now
show agreement in \neiii/\Hb, demonstrating  
insensitivity to shocks.  This is further supported by the data for DEM
301 (cross), which also falls in the locus expected for its \tstar.
Thus, \neiii/\Hb\ holds promise as a useful \tstar\ diagnostic for hot O stars.
Interestingly, the Galactic and LMC objects do not seem differentiated
as expected for their different metallicities.  Table~\ref{tabTcalib}
gives a preliminary empirical calibration for these \tstar\
diagnostics.  Further details can be found in Paper~I.  

\begin{table}[!b]
  \setlength{\tabcolsep}{2em} 
  \setlength{\tabnotewidth}{0.8\columnwidth} 
  \tablecols{3} 
  \begin{center}
    \caption{Empirical calibration of  \tstar\ 
	diagnostics\tabnotemark{a} }
    \label{tabTcalib}
    \begin{tabular}{lcc}\hline\hline
      Sp. type & log \neiii\lam3869/\Hb & log \etap \\ \hline
	WR &  $> -0.6$ & $< 0.2$ \\
	O3 -- O4 & --0.9 to --0.6 & \nodata \\
	O5 -- O8 & --1.5 to --0.9 & \nodata \\
	O9 and later & $<$ --1.5 & $> 1.0$ \\
      \hline\hline
      \tabnotetext{a}{For LMC metallicity.}
    \end{tabular}
  \end{center}
\end{table}

\section{Diagnostics of Metallicity:  $S234$}

\hii\ region metallicities are often estimated using the
``bright-line'' diagnostic of O/H:\ \ $R23 \equiv$ (\oii\ + \oiii)/\Hb\
(Pagel {\etal}1979).  Recently, a similar diagnostic has been
introduced for S:\ \  $S23 \equiv$ (\sii\ + \siii)/\Hb\ (V\'\i lchez \&
Esteban 1996).  Empirical calibrations for $S23$ have been presented
by Christensen {\etal}(1997) and D\'\i az \& P\'erez-Montero (2000),
who suggested that it would be less $U$-sensitive than $R23$ and would
also present a larger dynamic range.

\begin{figure}
\includegraphics[width=\textwidth]{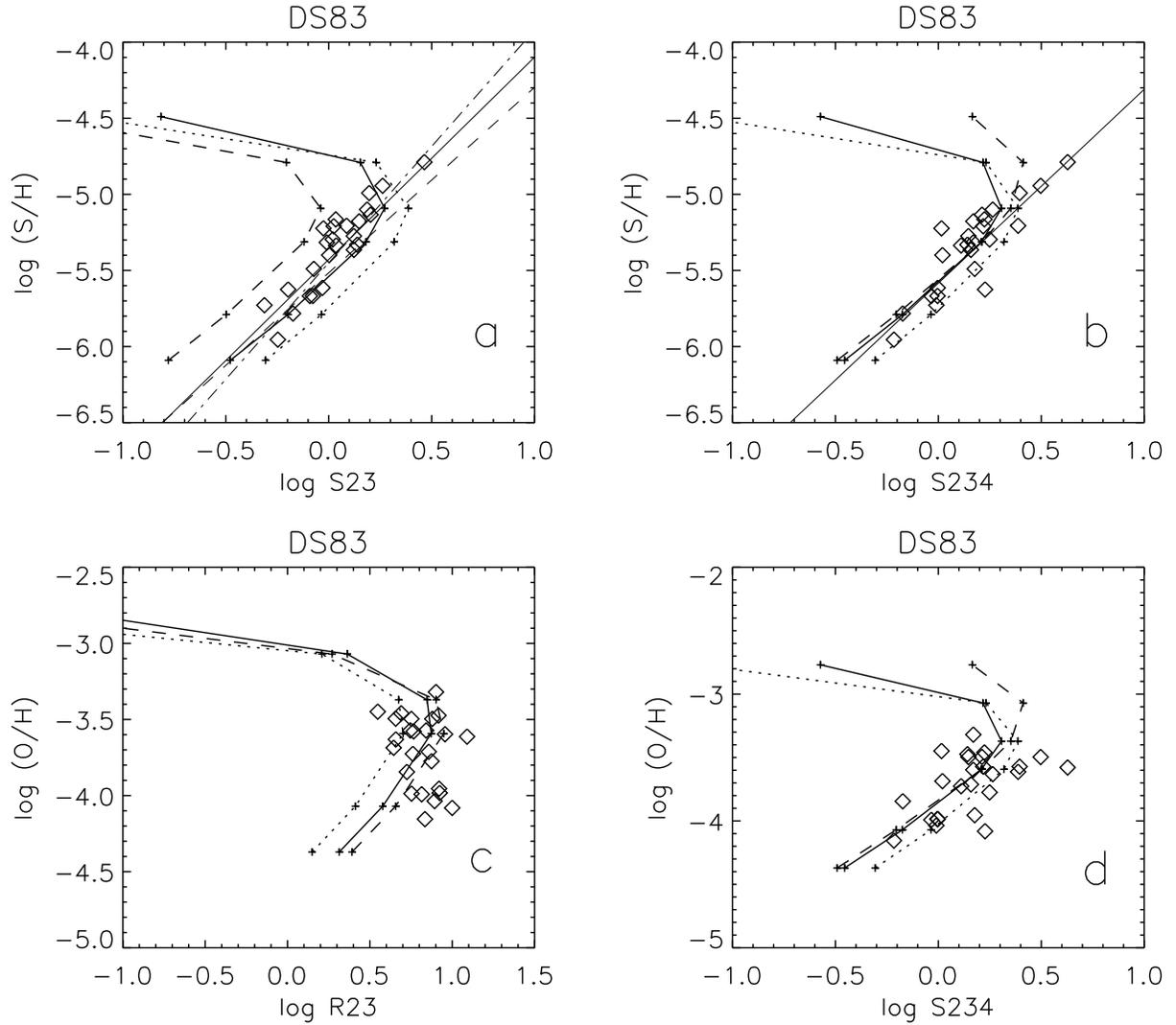}
\vspace*{-3.0in}
\caption{
Theoretical tracks for $a$) $S23$, $b$) $S234$, $c$) $R23$, and $d$)
log (O/H) vs. log $S234$, 
computed from models at 0.05, 0.1, 0.3, 0.5, 1.0, and 2.0 $\Zsol$.
Dashed, solid, and dotted lines correspond to $U = -2,\ -3,$ and --4,
respectively, and are overplotted with data from Dennefeld \& Stasi\'nska
(1983).  Straight lines are calibrations:  solid line is fitted to our
models for $Z \leq 0.5\Zsol$; dashed is from D\'\i az \&
P\'erez-Montero (2000); dash-dot from Christensen {\etal}(1997).  
\label{figDS}
}
\end{figure}

Figure~\ref{figDS} shows theoretical tracks for the abundance
diagnostics, using the same photoionization code and stellar
atmosphere models as before.  These are overplotted with nebular data
points from Dennefeld \& Stasi\'nska (1983), who obtained optical and
near-IR spectrophotometry of a large sample of nebulae, with
electron temperatures derived from the ratio $\lambda4363/\lambda5007$.
Our models show that $S23$ is actually more sensitive to $U$ than is
$R23$.  This 
results from the omission of \siv, which has roughly the same ionization
potential (35 eV) as \oiii.  The plot in Figure~\ref{figDS}$a$, of log(S/H)
vs. log $S23$, shows the large spread in the three tracks.  These
correspond to log $U = -2,\ -3$, and --4, with the first having the
lowest values of $S23$, as expected for a significant population of
\siv.  Note that this sequence of tracks is opposite to that for $R23$
(Figure~\ref{figDS}$c$), which more completely samples the important
ions.  

We therefore suggest an abundance diagnostic that includes \siv, to
better sample the relevant ions, and thereby improve the effectiveness
of the S diagnostic (Paper~II):
\begin{equation}\label{eqS234}
S234 \equiv \rm \bigl([S\thinspace II]\lambda6724 + 
	[S\thinspace III]\lambda\lambda9069,9532 +
	[S\thinspace IV]10.5\mu \bigr) / {\hb}
\end{equation}  
Although \siv$10.5\mu$ is a mid-IR line, its intensity can be
estimated from a simple relation between optical and near-IR lines,
for metallicities $Z \lta 0.5\Zsol$ (Paper~II): 
\begin{equation}\label{eqS4O3}
{\log{\frac{\rm [S\thinspace IV]10.5\mu}
	{\rm [S\thinspace III]\lambda\lambda9069,9532}} } = \break
	{-0.984 + 1.276\ \log{\frac{\rm [O\thinspace III]\lambda\lambda4959,5007}
	{\rm [O\thinspace II]\lambda3727}} }
\end{equation}
This relation provides a simple
way to estimate $S234$.  Figure~\ref{figDS}$b$ shows $S234$ for the
same dataset, with the intensity of \siv 10.5$\mu$ computed
from equation~\ref{eqS234}.  The theoretical tracks are dramatically
less sensitive to $U$ than for $S23$, or even $R23$
(Figure~\ref{figDS}$c$).  

For $S234$, a fit to our models gives a
theoretical calibration (Figure~\ref{figDS}$b$): 
\begin{equation}\label{calibS234}
\log {\rm(S/H)} = -5.58 + 1.27\ \log S234 \quad ,
\end{equation}
For $S23$, we have a similar theoretical calibration from the models
(Figure~\ref{figDS}$a$, straight solid line): 
\begin{equation}\label{calibS23}
\log ({\rm S/H}) = -5.43 + 1.33\ \log S23 \quad .
\end{equation}
The dashed and dot-dashed lines in Figure~\ref{figDS}$a$ show
calibrations from Christensen 
{\etal}(1997) and D\'\i az \& P\'erez-Montero (2000) for comparison.
Figure~\ref{figDS}$d$ shows that caution is necessary in inferring
O abundances from the S diagnostics, since apparently there is
significant scatter in S/O.  Figure~\ref{figDS} also shows
that all of these calibrations are reliable only for $Z \lta 0.5 \Zsol$.

\acknowledgements
We thank Angeles D\'\i az, Rob Kennicutt, and Fabio Bresolin for
providing access to their data in advance of publication.  
MSO gratefully acknowledges financial support from the conference
organizers that enabled her attendance at this celebration.



\begin{thebibliography}

\bibitem{} Christensen, T., Petersen, L, \& Gammelgaard, P., 1997,
	AA, 322, 41
\bibitem{} Dennefeld, M. \& Stasi\'nska, G., 1983, AA, 118, 234
\bibitem{} D\'\i az, A. I. \& P\'erez-Montero, E., 2000, MNRAS, 312, 130
\bibitem{} Kennicutt, R. C., Bresolin, F., French, H., \& Martin, P.,
	2000, ApJ, 537, 589
\bibitem{} Oey, M. S., Dopita, M. A., Shields, J. C., \& Smith,
	R. C. 2000, ApJS, 128, 511 (Paper~I)
\bibitem{} Oey, M. S. \& Shields, J. C. 2000, ApJ, 539, 687 (Paper~II)
\bibitem{} Pagel, B. E. J., Edmunds, M. G., Blackwell, D. E., Chun,
	M. S., \& Smith, G., 1979, MNRAS, 189, 95
\bibitem{} Peimbert, M., 1967, ApJ, 150, 825
\bibitem{} Peimbert, M., Sarmiento, A., \& Fierro, J., 1991, PASP,
	103, 815
\bibitem{} Schaerer, D. \& de Koter, A., 1997, AA, 322, 598
\bibitem{} Schmutz, W., Leitherer, C., \& Gruenwald, R., 1992, PASP,
	104, 1164
\bibitem{} Sutherland, R. S. \& Dopita, M. A. 1993, ApJS, 88, 253
\bibitem{} V\'\i lchez, J. M. \& Esteban, C. 1996, MNRAS, 290, 265
\bibitem{} V\'\i lchez, J. M. \& Pagel, B. E. J., 1988, MNRAS, 231, 257



\end{thebibliography}
\end{document}